\documentclass[multphys,vecphys]{svmult}

\usepackage{url}             
\usepackage{makeidx}         
\usepackage{graphicx}        
\usepackage{multicol}        
\usepackage[bottom]{footmisc}

\makeindex             

\newcommand{\hlmicron}{\hbox{$\mu$m}}
\newcommand{\hlla}{\mathrel{\hbox{\rlap{\hbox{\lower4pt\hbox{$\sim$}}}\hbox{$<$}}}}

\begin{document}    

\title*{
{\em Spitzer\/} 4.5~$\mu$m Luminosity-Metallicity and Mass-Metallicity
Relations for Nearby Dwarf Irregular Galaxies
}

\titlerunning{$L$--$Z$ and $M$--$Z$ Relations for Dwarf Irregulars}

\author{
Henry Lee\inst{1}, E. D. Skillman\inst{1}, J. M. Cannon\inst{2},
D. C. Jackson\inst{1}, R. D. Gehrz\inst{1}, E. Polomski\inst{1},
\and C. E. Woodward\inst{1}
}

\authorrunning{H.~Lee et al.}

\institute{
Dept.\ of Astronomy, Univ.\ of Minnesota, 116 Church St. S.E.,
Minneapolis, MN 55455 USA
(\texttt{hlee, skillman, djackson, gehrz, elwood, chelsea@astro.umn.edu})
\and 
Max-Planck-Institut f\"ur Astronomie, K\"onigstuhl 17,
D-69117 Heidelberg, Germany
(\texttt{cannon@mpia.de})
}
\maketitle

\begin{abstract}
For a sample of 25 dwarf irregular galaxies with distances 
$D \hlla 5$~Mpc and measured oxygen abundances, we present results
derived from galaxy luminosities at 4.5~\hlmicron\ and stellar masses
from near-infrared imaging with IRAC on the {\em Spitzer Space Telescope}.
We have constructed the appropriate luminosity-metallicity ($L$--$Z$)
and mass-metallicity ($M$--$Z$) relations, and compared these
relations with their corresponding relations from the Sloan Digital
Sky Survey (SDSS).
We obtain the following results.
1. The dispersion in the near-infrared $L$--$Z$ relation is reduced
with respect to the dispersion in the optical $L$--$Z$ relation, which
agrees with expectations for reduced variations of stellar
mass-to-light ratios at longer wavelengths compared to optical
wavelengths.
2. The dispersion in the optical $L$--$Z$ relation is similar over
approximately 11~mag in optical luminosity.
3. With our constructed $M$--$Z$ relation, we have extended the SDSS
$M$--$Z$ relation to lower masses by about 2.5~dex in stellar mass.
4. The dispersion in the $M$--$Z$ relation appears to be comparable
over a range of 5.5~dex in stellar mass.
\end{abstract}

\section{Near-Infrared Luminosity-Metallicity Relation}

The luminosity-metallicity ($L$--$Z$) relation for nearby dwarf irregular
galaxies has been studied traditionally at optical wavelengths
(e.g., \cite{lee:lequeux79,lee:skh89,lee:rm95,lee:lee03,lee:vanzee06}).
However, the dispersion in the optical $L$--$Z$ relation is affected
by variations in stellar mass-to-light ratios, which are caused by 
variations in the current star formation rate among galaxies.
To minimize the effects of these variations, we determine the $L$--$Z$
relation at near-infrared wavelengths where stellar populations
dominate the emission.
The sensitivity of the {\em Spitzer Space Telescope} provides
an excellent opportunity to observe the total near-infrared emission
from the stellar populations in nearby dwarf galaxies.
The present results are based on observations taken with IRAC in
channel 2 (4.5~\hlmicron); we assume that the contribution from warm
and/or small dust grains at 8~\hlmicron\ is negligible in dwarf
galaxies \cite{lee:jackson06}.

\begin{figure}
\centering
\includegraphics[height=10cm,angle=-90]{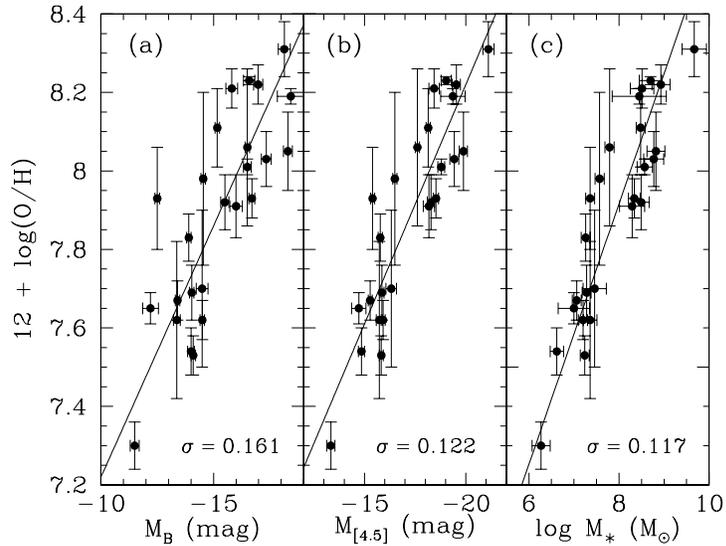}
\caption{
Panel~(a): optical ($B$) luminosity-metallicity relation with
dispersion $\sigma$ = 0.16~dex.
Panel~(b): near-infrared (4.5~\hlmicron) luminosity-metallicity relation
with dispersion $\sigma$ = 0.12~dex.
Panel~(c): stellar mass-metallicity relation with dispersion
$\sigma$ = 0.12~dex.
In each panel, the solid line represents a linear least-squares fit to
the data.
}
\label{lee:fig1}
\end{figure}

We have measured 4.5~\hlmicron\ luminosities for 25 galaxies
taken from GTO program 128 (P.I. R.~D. Gehrz) and the {\em Spitzer\/}
archive.
Located within the Local Group and other nearby groups,
these galaxies have measured distances ($D \hlla 5$ Mpc) and
oxygen abundances.
The optical and near-infrared $L$--$Z$ relations \cite{lee:irlz} are
plotted in Figs.~\ref{lee:fig1}a and b, respectively.
The dispersion in the $L$-$Z$ relation is reduced at near-infrared
wavelengths compared to the dispersion in the optical relation.
By comparison with the optical $L$--$Z$ relation for more massive
galaxies \cite{lee:tremonti04} from the Sloan Digital Sky Survey (SDSS),
it appears that the dispersion in the optical $L$--$Z$ relation is
similar ($\simeq$ 0.16~dex) over 11 magnitudes in optical luminosity.

\section{Stellar Mass-Metallicity Relation}

With measured luminosities at 4.5~\hlmicron, we have derived
stellar masses for our sample of dwarf irregular galaxies.
We have used the models from Bell \& de Jong \cite{lee:belldejong01}
to determine stellar mass-to-light ratios as functions of $B\!-\!K$
color, and we have applied a correction for nonzero $K\!-\![4.5]$
color in late-type dwarf galaxies (e.g., \cite{lee:pahre04}).
We have also adjusted the derived stellar masses 
to the Kroupa \cite{lee:kroupa01} stellar initial mass function;
additional details of the derivation are provided in \cite{lee:irlz}.

\begin{figure}
\centering
\includegraphics[height=10cm,angle=-90]{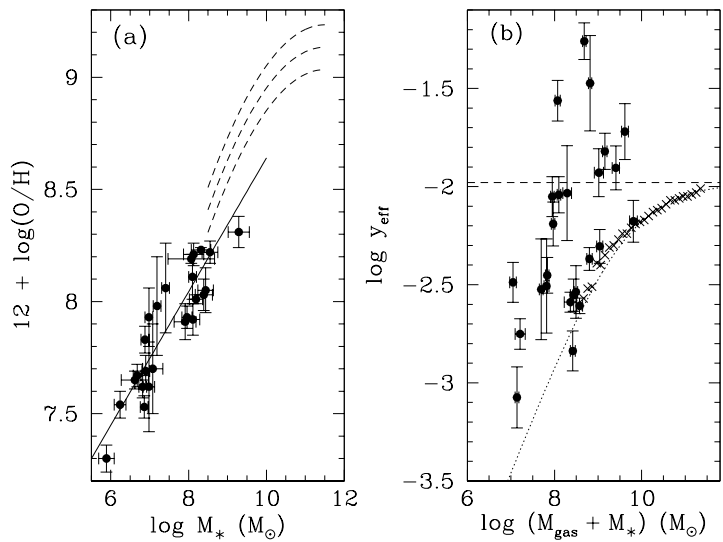}
\caption{
Panel~(a): Stellar mass-metallicity relation over 5.5~dex
in stellar mass.
The filled circles and solid line represent the present sample
of dwarfs and the best fit, respectively.
The relation from the SDSS and the $\sim$ 0.10~dex curves
from \cite{lee:tremonti04} are shown as dashed lines.
Panel~(b): Effective yield versus total baryonic mass.
The crosses represent the median of the SDSS data in mass bins
of 0.1~dex \cite{lee:tremonti04}.
An empirical fit is shown as a dotted line, and the asymptotic
yield ($y_{\rm eff}$ = 0.0104) is shown as a horizontal dashed line.
}
\label{lee:fig2}
\end{figure}

The stellar mass-metallicity ($M$--$Z$) relation \cite{lee:irlz} is 
shown in Fig.~\ref{lee:fig1}c.
The $M$--$Z$ relations for dwarf galaxies and for massive
galaxies from the SDSS are shown in Fig.~\ref{lee:fig2}a.
With the present sample of dwarf galaxies, we have extended the 
SDSS $M$--$Z$ relation to lower stellar mass by 2.5~decades.
We find that the dispersion in the $M$--$Z$ relation is 
comparable ($\approx$ 0.10--0.12 dex) over a range of 5.5 decades in
stellar mass, although we have not performed a homogeneous treatment of 
gas-phase metallicities and stellar masses for the SDSS sample; 
see \cite{lee:irlz} and \cite{lee:tremonti04} for details.

We have plotted the effective yield as a function of total baryonic
mass in Fig.~\ref{lee:fig2}b.
The simple closed-box model of chemical evolution (see, e.g.,
\cite{lee:ss72}) predicts that the effective yield is equal to the true
yield for all masses.
However, that the effective yield decreases at lower baryonic mass is
commonly interpreted as a signature of either outflow or dilution from
the infall of metal-poor gas.
We find that the present sample of dwarf galaxies exhibits a much
larger variation in the effective yield at a given total baryonic mass.
The large variation in the effective yield is difficult to explain if
galaxy outflows are dominant in low-mass dwarf galaxies;
see \cite{lee:tremonti04} for a countering view. 

\section{Stellar Iron Abundances vs. Near-Infrared Luminosity}

\begin{figure}
\centering
\includegraphics[height=10cm,angle=-90]{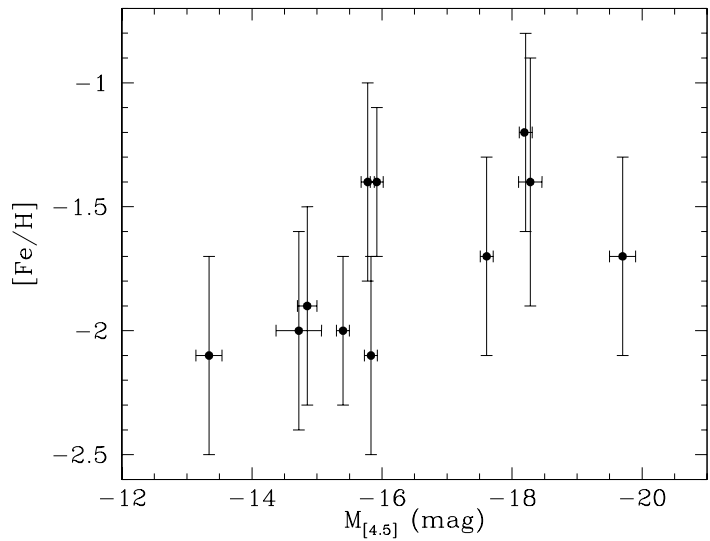}
\caption{
Stellar photometric iron abundances vs. absolute magnitude at 
4.5~\hlmicron.
The iron abundances were taken from \cite{lee:ggh03}.
}
\label{lee:fig3}
\end{figure}

Stellar masses for our sample of dwarf galaxies have been derived from
near-infrared luminosities under the assumption that the emission is
dominated by the populations of older stars.
Iron abundances are a good tracer of the chemical evolution for these
stars integrated over the entire history of past star formation.
We have constructed an $L$--$Z$ relation using photometric stellar
iron abundances from the literature to examine if the correlation is
comparable to the relation constructed using gas-phase metallicities.
The iron abundance-luminosity relation is shown in
Fig.~\ref{lee:fig3}.
Unfortunately, the photometric iron abundances have large
uncertainties or spreads.
Nevertheless, there appears to be a trend between the
mean photometric iron stellar abundance and the measured
galaxy luminosity at 4.5~\hlmicron\ in the same sense as
that found in Fig.~\ref{lee:fig1}.

\section{Final Remarks}

The challenge is explaining the relatively uniform scatter in both 
the $L$--$Z$ and $M$--$Z$ relations.
Additional near-infrared imaging and spectroscopy for a large number
of dwarf galaxies within the Local Volume can test whether the scatter
in the $L$--$Z$ and $M$--$Z$ relations holds over a large range in
luminosity and mass. 
The growing interest in determining the redshift-evolution of the
$L$--$Z$ and $M$--$Z$ relations provides good impetus to 
exploring galaxy formation models that incorporate varying degrees of
galaxy outflows and can predict the slope and the scatter in the
$L$--$Z$ and $M$--$Z$ relations over large dynamic range.

%
\index{Dwarf irregular galaxies}
\index{IRAC}
\index{Local Volume}
\index{Mass-metallicity}
\index{Spitzer}
\printindex
\end{document}